\documentclass{ws-p10x7}

\usepackage{amssymb,amsmath}

\begin{document}

\title{High Energy Heavy Ion Collisions: The Physics of Super-dense Matter}

\author{Barbara V. Jacak}

\address{Department of Physics and Astronomy, SUNY Stony Brook, Stony Brook,
NY, 11794,
USA\\E-mail: jacak@nuclear.physics.sunysb.edu}

\twocolumn[\maketitle\abstract{
I review experimental results from ultrarelativistic heavy ion collisions. 
Signals of new physics and observables reflecting the underlying 
collision dynamics are presented, and the evidence for new physics discussed. 
Measurements of higher energy collisions at RHIC are described, and I give
some of the very first results. }]

\section{Introduction}

High energy heavy ion collisions aim to recreate the 
conditions which existed a few microseconds following the big bang,
and determine the properties of this super-dense matter. 
The density of produced hadrons is very high; at energy
densities of 2-3 GeV/fm$^3$, the inter-hadron distance is smaller 
than the size of the hadrons themselves. Interactions 
among hadrons under such conditions are unlikely to be the 
same as in the familiar dilute hadron gas. QCD predicts 
that at sufficiently high energy density and temperature, the 
vacuum ``melts" into numerous $q \overline q$ pairs.

Such matter is expected to leave the realm where quarks and 
gluons are confined in colorless hadrons, and form, instead, a quark-gluon 
plasma. The experiments explore two fundamental puzzles of QCD,
namely the confinement of quarks and gluons into hadrons, and the
breaking of chiral symmetry which produces mass of the constituent
quarks. We aim to study experimentally the nature of deconfined matter, 
investigate the confinement phase transition, and
determine its temperature. The chiral transition is expected to occur 
under similar conditions. 
Use of the heaviest ions maximizes the volume and lifetime of matter 
at high energy density, enhancing signals of new physics. Understanding 
the background from high energy hadronic collisions, as well as the
underlying dynamics and nuclear structure is accomplished via p+p and 
p+nucleus collisions in the same detectors.

Solutions of QCD on the lattice have been used to estimate the 
energy density required for deconfinement.\cite{karsch} 
In calculations with three massless
quark flavors, a rapid change in the energy density occurs at a 
critical temperature of 170 $\pm$ 10 MeV. The energy density at
which the system is fully in the new phase is approximately 3 
GeV/fm$^3$. With 2 massless and one strange quark, the critical 
energy density is 15\% lower. Studies have shown that the mass of 
the $<q\overline q>$ condensate falls to zero, signifying 
restoration of chiral symmetry, at about the same temperature. 

\subsection{Early stage and evolution of the collision}

Experimental access to information about the high energy density 
phase is complicated by the subsequent expansion, cooling and 
re-hadronization of the matter. Theoretically, however, one
may consider several separate stages of a heavy ion collision.
Interpenetration and initial nucleon-nucleon collisions 
are complete in less than 1 fm/c. This is accompanied by multiple 
parton collisions leading, probably, to local
thermal equilibration. The hot, dense matter expands longitudinally
and transversely, cooling until the quarks re-hadronize. The
hadrons continue to interact among themselves until the system
is sufficiently dilute that their mean free path exceeds the size
of the collision zone. At this point, hadronic interactions cease
and the system ``freezes out''.

Elementary nucleon-nucleon
collisions have long been studied, and a wealth of data on
$p-p$ and $\overline p-p$ collisions are in the literature.
Quantitative understanding of the initial parton production 
in heavy ion collisions requires starting with the nucleon 
quark and gluon structure functions, which are now rather 
well known from deep inelastic e-p and from p-nucleus 
experiments.\cite{stfunc} A steep rise of $F_2$, the quark
structure function, was discovered\cite{stfunc} toward low $x$ 
for $Q^2 \ge$ 2 GeV$^2$. This rise is understood to indicate
the dominance of gluons, and implies very large numbers of
gluon-gluon interactions when two nuclei collide at high 
energy. H1 at HERA has unfolded the gluon distribution from
their data,\cite{h1} and finds a steep rise at small $x$.
Following this observation, we may expect significant enhancement
of gluon fusion processes, such as charm production, for example,
in heavy ion collisions. The $x$ and $Q^2$ regions of interest at 
RHIC are $x \ge$ 0.01 and $Q^2 \approx$ 10-20 GeV$^2$.

It has been observed, however, that structure functions of quarks
in nuclei differ from those of nucleons. There is a depletion 
of quarks in the small-x region, known as ``nuclear shadowing'';
this effect is expected in gluon distributions also. Shadowing 
is usually attributed to parton fusion preceding the hard scattering 
which probes the parton distribution. As the overcrowding at
small $x$ is larger in nuclei than in individual nucleons, 
saturation should be more evident for heavy nuclei, causing 
shadowing to strengthen with nucleon number. This is indeed the
case, and measurements show a 20\% modification is heavy nuclei
at $x = 0.03$. For Au nuclei, the shadowing in this $x$ region
should be a 30 \% effect. Shadowing may reduce the gluon momentum
requiring corresponding enhancement in the large x region if the
momentum fraction of gluons is to be conserved. Such ``anti-shadowing''
has been predicted by Eskola and co-workers, using a DGLAP 
evolution.\cite{eskola} The total number of charged particles produced
in a heavy ion collision is sensitive to the magnitude of shadowing
and anti-shadowing effects, and can be used to constrain the evolution
calculations.

Even with nuclear shadowing, the density of partons after the initial
hard nuclear collision is truly enormous, leading us to expect a large
amount of multiple parton scattering. Such multiple scattering is already
visible in proton-nucleus collisions as the Cronin effect, which hardens
the pion $p_T$ spectra above 1.5 GeV/c. The higher parton density in 
nucleus-nucleus collisions should drive the system toward thermal equilibrium
by thermalizing mini-jets and increasing the multiplicity of soft particles.
Indeed, parton cascade descriptions
of the collision dynamics predict that equilibration 
among the partonic degrees of freedom happens within 
0.3-1 fm/c in collisions at $\sqrt{s}$ = 200 GeV/A.\cite{geiger} 

The dense medium should affect fast quarks traversing it, and in fact
a medium-induced energy loss of partons is expected. As first predicted
by Gyulassy and coworkers,\cite{gyuwang} and Baier, Dokshitzer, Mueller, 
Peigne and Schiff,\cite{bdmps} the energy loss of a fast quark increases 
with the
density of the medium, due to an accumulation of the transverse momentum
transferred. The energy loss dE/dx may exceed 1 GeV/fm, and BDMPS calculated
that it could reach $ 3 /times (L/10fm)$ GeV/fm at $T=$250 MeV, 
where $L$ is the path
length through the dense medium. Experimental measurements of this energy
loss will thus reflect the density of the medium early in the collision.

The quark gluon plasma expands and cools,
whereupon the system hadronizes. The deconfined and mixed phases are
expected to last approximately 3 fm/c, after which the system becomes 
a dense, interacting hadron gas. Expansion continues, and the system
finally becomes sufficiently dilute that the hadrons cease to interact
approximately 10 fm/c after the start of the collision.\cite{geiger} 

Of course, these values depend strongly upon the assumptions in the
models, and the boundaries between phases are not sharp in either time
or space. A major experimental challenge is to determine the 
timescales, along with the duration of hadron emission following
freezeout. The expansion velocity is accessible via interferomety; 
scaling longitudinal expansion ($\beta \approx 1$)
along with radial expansion at 
approximately half the longitudinal velocity have been 
observed.\cite{NA49HBT}

\subsection{Predicted signals of quark gluon plasma}

A number of key predictions for quark gluon plasma signatures were made
prior to experiments at CERN and Brookhaven. Color screening by a quark gluon 
plasma was predicted to suppress bound $c \overline c$ pairs, resulting in 
decreased $J/\psi$, $\psi^\prime$  and  $\chi_c$ production.\cite{matsatz} 
Observation of this effect is subject to understanding final 
state interactions of the charmed mesons with nucleons and co-moving 
hadrons, which break up the bound state.

Rafelski and Mueller predicted
in 1982 that production of strange hadrons should be enhanced by formation
of quark gluon plasma.\cite{rafmueller} 
The rate of gluon-gluon collisions rises in a hot gluon gas,
thereby increasing the cross section
of gluon fusion processes and production of strange and charmed quarks.
An important hadronic background to this measurement is 
associated production of strange particles in the dense hadron gas,
primarily via $\pi + \rm N \rightarrow \rm K + \Lambda$. Strangeness
exchange also complicates the picture. 

Thermal electromagnetic radiation reflects the initial temperature
of the system, via quark-antiquark annihilation to virtual photons which
decay to lepton pairs, and via quark-gluon Compton scattering. The rate, 
proportional to $T^4$, should be dominated by the initial
temperature, $T_{init}$; the shape
of the spectrum will reflect this maximum temperature. Measurements are 
difficult because of the large lepton and photon backgrounds from
hadron decays, hadronic bremsstrahlung, D meson decays and Drell-Yan
pairs.

Possible observable effects of chiral symmetry restoration
include modification of meson masses\cite{hadmass} 
(visible through their leptonic decays) and formation of disoriented 
chiral condensate domains. Such a condensate should result 
in modified ratios of charged to neutral pions and enhanced production 
of soft pions with $p_T \le$ 100 MeV/c.\cite{dcc}

The predicted large energy loss of quarks traversing a very dense 
medium would result in ``quenching'' of jets,\cite{gyuwang,bdmps}
which can be observed experimentally via the spectrum of 
hadrons at high $p_T$. Since such hadrons are dominantly the 
leading particles in jet fragmentation, their spectrum reflects 
the spectrum of quarks exiting the medium. This observable
reflects the density of the medium, rather than its confinement 
properties, but experimental evidence for existence of a superdense
medium would be most exciting.

\subsection{Experimental observables}

The charged particle multiplicities in high energy heavy ion collisions 
are enormous. At SPS energy of $\sqrt{s}$ = 18-20 GeV/A,
there are more than 1000 hadrons produced, while at RHIC the number is
closer to 10000.

Experimentally accessible observables fall into two classes. 
The first characterize the system formed and 
ascertain that the conditions warrant a search for new 
physics. These observables are primarily hadronic and reflect the 
system late in the collision. Detailed analysis of hadrons also 
yields dynamical information about the collision, allowing one to 
extrapolate the hadronic final state back to the hottest, densest 
time when quark-gluon plasma may have existed.\cite{hbtreview}

The second class of observables comprises signals of new physics. 
Lepton pairs and photons ($\it i.e.$ virtual and real photons) 
decouple from the system early, and emerge undisturbed 
by the surrounding hadronic matter. Consequently, their  
distributions are dominated by the early time in the collision, 
and the rate reflects the initial temperature. Strangeness
production can be detected via $K, \Lambda$ and other hadrons 
containing strange quarks; multistrange anti-baryons are particularly 
promising indicators of strangeness enhancement, as it is 
difficult to affect their production via hadronic 
means.\cite{rafelski} In a high 
temperature plasma with many gluons, the gluon fusion reaction 
$g + g \rightarrow c \overline c$ should be 
important.\cite{charmcasc,charmtherm,geiger} Measurement of
semileptonic, or perhaps 
even fully reconstructed, decays of charmed mesons would indicate 
whether charm production reflects enhanced gluon fusion.

\section{Results for $\sqrt s \le$ 20 GeV/nucleon}

\subsection{Energy Density}

Before searching for evidence of deconfinement, we must determine whether
appropriate values of energy density are in fact reached. Estimating
the energy density 1 fm/c into the collision from measured quantities
requires some assumptions. However, this can be done from measured production
of energy transverse to the beam direction, $E_T = \Sigma E \sin\theta$. $E_T$
reflects the randomization of the incoming longitudinal energy of the beam.
For collisions undergoing scaling longitudinal expansion, the energy density
may be estimated via $\epsilon = \rm d E_T/\rm d \eta \times 1/\rm{volume}$. The
volume is given by the cross sectional area of the nucleus involved, and a
length defined by the formation of particles, $\tau$, generally
taken to be 1 fm, though this is likely $\sqrt{s}$ dependent, becoming smaller
for high energy collisions. 

Selecting as central collisions, the few percent of the total cross
section producing the largest multiplicity,
one may estimate the relevant nuclear area to be $\approx 90\%$ 
of the total. Using the $E_T$ value of 450 GeV
at this point, as measured in Pb + Pb collisions by the NA49 
collaboration \cite{na49et}, and $R = 1.2 A^{1/3} (0.9)$, with 
$V = (\pi R^2) \tau$, the energy density, $\epsilon$, is found to be 
$\approx$ 3.2 GeV/fm$^3$. This is sufficiently high compared to the predicted
transition point, to encourage searching for signals of new physics.

\subsection{Color screening }

As charmed quarks are produced in the initial hard collisions, they 
traverse the dense matter and therefore probe its properties. The screening
length is directly related to the temperature and energy density, so $c 
\overline c$ bound states of different radius will be screened
under different conditions. The $J/\psi$, with radius of 0.29 fm and 
binding energy of approximately 650 MeV should be more stable than the
$\psi^\prime$ with binding energy of 60 MeV and nearly twice the radius.

Suppression of $J/\psi$ 
production has been observed by NA50.\cite{jpsi} In light systems, the 
suppression is consistent with expectations from initial and final 
state effects on production and binding of the $c \overline c$ 
pairs. However, in Pb + Pb collisions, the suppression is 25\% 
greater than that expected from conventional 
processes. The anomalous suppression sets in for semi-peripheral 
Au + Au collisions, and increases in strength with the volume of 
the excited system. 

\begin{figure}
\epsfxsize=6.5cm
\epsfbox{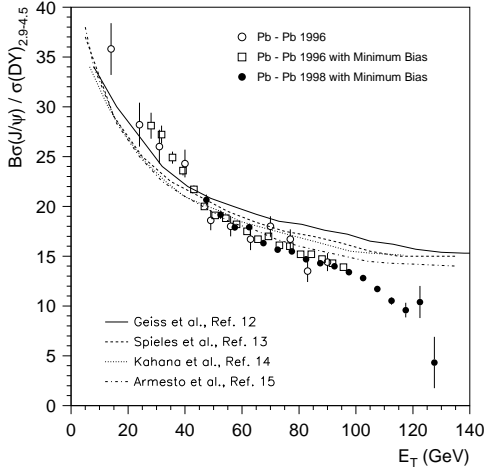}
\caption{Comparison of NA50 measurement of the ratio of $J/\psi$
and Drell-Yan cross section as a function of $E_T$ (i.e. centrality)
in Pb + Pb collisions with several
conventional descriptions of $J/\psi$ suppression \protect\cite{na50}.}
\label{fig:na50}
\end{figure}

The observed suppression has been compared with hadronic models
as a function of collision centrality (determined by measurement of
transverse energy).\cite{na50}
The measured $J/\psi$ to Drell-Yan ratio decreases in more central
collisions, and by $E_T \approx$ 100 GeV has fallen well below
models which assume charmonium states are absorbed by interactions
with comoving hadrons.\cite{na50}
Figure 1 shows the measured ratio of $J/\psi$ to Drell-Yan in 158 
GeV/A
Pb + Pb collisions, as a function of $E_T$; high $E_T$ corresponds 
to central Pb + Pb collisions.\cite{na50} The points are data,
while the curves show $J/\psi$ production in models which assume
that the charmonium states are absorbed by interactions with
comoving hadrons.\cite{models}
Discussion continues within the community regarding 
discontinuities and thresholds, and a possible
second drop of the $J/\psi$ production at $E_T \ge$ 110 GeV. 
Nevertheless, the data very clearly deviate from the hadronic
models for central collisions.

\subsection{Strangeness enhancement}

Strangeness production in a heavy ion collision may not be subject
to the well-known ôstrangeness suppressionö observed in elementary 
nucleon collisions. This can be easily understood from simple energy
level considerations. If the energy levels of u and d quarks are empty,
it is energetically favorable to produce these light quarks, since the 
s quark levels have an energy gap of twice the strange quark mass. In
nucleus-nucleus collisions, however, the dense matter causes the lowest
u and d quark levels to be filled, resulting in a relative enhancement
of strange quark number over p-p collisions.

Enhancement of several species of strange hadrons has been observed.
Kaon and $\Lambda$ enhancement may be expected in a dense hadron 
gas,\cite{rqmdk} as a result of associated production in hadron 
multiple scattering, but the excess production of 
$\overline \Lambda$, $\Xi$ and $\Omega$ observed by WA97 
is not easily explained without 
new physics.\cite{wa85,wa97} It is particularly remarkable that the
enhancement compared to p-nucleus collisions of multiply strange 
antibaryons increases with the number of strange quarks.
In order to achieve such production rates via hadronic equilibration,
a dense hadron gas would need to live for about 100 fm/c. 
Given the measured expansion velocities, such a lifetime is ruled out.

\subsection{Thermal radiation}

Measurable rates of thermal dileptons and thermal photons were
predicted by Shuryak and others.\cite{thermal}
The rate is proportional to the fourth power of the temperature, and is 
thus dominated by the initial (highest) temperature of
the system. The distribution of the radiation should have an exponential
shape $$\propto e^{-M/T},$$ and the rate should depend on the square 
of the particle multplicity. Consequently, detection of thermal radiation 
depends strongly on $T_{init}$ achieved in the collisions. Complicating both 
measurement and interpretation, and boosting the photon emission rate, 
is partonic bremsstrahlung, which also contributes.\cite{aurenche}

It is important to note that a hot hadron gas will radiate thermal 
photons and dileptons as well. Consequently, thermal radiation does
not indicate the phase of the matter, but reflects its highest temperature.
Experimentally, the goal is to measure the presence of real photons or
dileptons, beyond those from hadronic decays, and from the yield and
distributions extract $T_{init}$. This value is then to be compared 
to the expected transition temperature.

WA98 observed direct photons beyond contributions
from hadronic decays in central Pb + Pb collisions at 158 
GeV/nucleon at $p_T \gtrsim 2$ GeV/c.\cite{wa98photons}
A decomposition of the
excess photons does not clearly show whether the $p_T$
distribution differs from photons in p-p and p-nucleus collisions, but
the enhanced yield appears consistent with $T_{init} > T_C$.
\cite{wa98photons,srivastava}

The spectrum of lepton pairs below the $\rho$ meson mass is of 
considerable interest. Though in p+Be and p+Au collisions the
invariant mass distribution of electron-positron pairs below 1 GeV
is well described by hadronic decays, a clear excess is observed
in S + Au and Pb + Pb collisions by CERES.\cite{ceres} It is 
difficult to reproduce the observed distribution without allowing 
the mass of the $\rho$ to change or invoking a tremendous amount of 
collision broadening, which opens new phase space and effectively 
lowers the $\rho$ mass.\cite{brown} Such observations 
may indicate partial chiral symmetry restoration in the dense 
matter created in collisions at the SPS.

It may be that the low mass dileptons arise 
from thermal radiation.\cite{kaempfer} Excess dileptons in the 
intermediate mass range between 1 and 3 GeV have also been observed.
The intermediate mass lepton pair cross section can be explained if 
thermal radiation is added to Drell-Yan and charm decay 
sources.\cite{rapp} Both sets of data imply initial temperatures
in the range 170-200 MeV. Both calculations make use of parton/hadron
duality in the dense system to predict the thermal radiation from
simple $q \overline q$ annihilation rates, integrating over the
time evolution of the collision. Such explanations do not prove
that the observed dileptons are thermal in origin, but indicate
that the spectra are consistent with an initial temperature near 
or above the predicted phase transition temperature.

\subsection{Evolution of the hadronic phase}

The high density of produced particles should create high 
pressure in the collision, leading to rapid expansion. The
expansion velocity can be extracted by combining 
measurements of single particle $m_T$ distributions with two 
particle correlations;\cite{hbtreview} 
requiring a simultaneous fit of both distributions to 
disentangle flow from thermal motion.

The inverse slopes of single particle spectra are given 
by $$T \approx T_{freezeout} + \frac{1}{2} m_0 <v_T>^2$$
where $T_{freezeout}$ is the temperature at which the hadronic system
decouples (i.e. hadron collisions cease) and $<v_T>$ is the average
radial expansion velocity.

The two particle correlations measure the size of the region 
of hadron homogeneity (i.e. full information transport) at freezeout. 
Position-momentum correlations from expansion case this to be
smaller than the entire hadron gas volume.
Large statistics are needed for 3-dimensional analysis of the 
correlation functions, binned in $m_T$ of the particles. The results 
of such analyses follow approximately 
\begin{displaymath}
R_T^2 = \frac{R^2}{1 + \xi \frac{m_T}{T_{freezeout}}<v_T^2>}
\end{displaymath}
The freezeout temperature is approximately
100 MeV and the average radial expansion velocity ~0.5 c. The
data indicate that the system expands by a factor of 3 radially
while undergoing a scaling expansion longitudinally. Back-extrapolation
from the freezeout conditions, combined with the measured transverse
energy yields energy densities of 2-3 GeV/fm$^3$.

\begin{figure}
\epsfxsize=6.5cm
\epsfbox{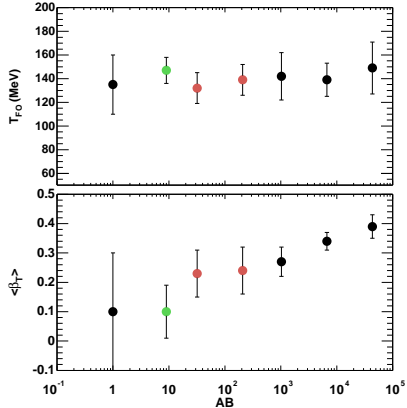}
\caption{Dependence of freezeout temperature and expansion velocity
extracted from $\pi$, K, and p spectra below $m_T$ = 1 GeV on colliding
system size \protect\cite{jbh}.}
\label{fig:jane}
\end{figure}

If the radial expansion velocity, $v_T$, indicates 
pressure created in the collision, $v_T$ should increase
with the number of produced particles. Burward-Hoy\cite{jbh} performed 
a global study of $v_T$ with system size by analyzing single
particle spectra below $m_T = 1$ GeV/c; correlation functions are only 
available for a small subset of projectile-target combinations. The 
radial expansion velocity indeed
increases with system size, as can be seen in Figure 2, which
shows the extracted freezeout temperature and $v_T$ as a function of 
the number of possible nucleon-nucleon collisions 
($A_{projectile} \times A_{target}$). 
The apparent freezeout temperature is 140 MeV, approximately independent 
of system size. Radial expansion in the large 
colliding systems boosts the particles, thus soft physics processes reach
larger $p_T$ than in elementary collisions. Consequently, observations
of hard scattering will require higher $p_T$. Burward-Hoy extrapolated
the soft spectrum using the $T_{FO}$ and $v_T$ parameters and found that
for Pb + Pb at CERN, hard scattering is only a partial contribution
to the spectrum below $p_T \approx$ 5 GeV/c.\cite{jbh}

\subsection{Quark energy loss}

\begin{figure}
\epsfxsize120pt
\figurebox{}{}{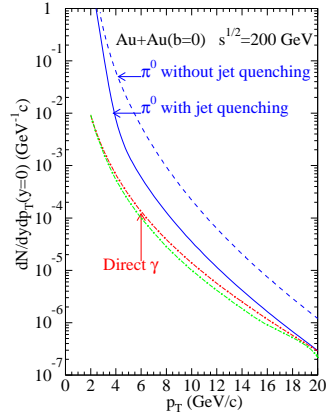}
\caption{$p_T$ distribution for $\pi^0$ with and without parton energy 
loss as compared to direct photons in central Au + Au at $\sqrt{s}$ = 200 
GeV/nucleon. $dE/dx$ = 1 GeV/fm was used \protect \cite{wangjetq}.}
\label{fig:jetq}
\end{figure}

Figure 3 shows predictions by X.N. Wang of the effect of quark energy loss
on the single particle $p_T$ spectrum.\cite{wangjetq} At sufficiently high 
$p_T$, where the
spectrum is dominated by leading particles from jet fragmentation, energy loss
or jet quenching will decrease the yield of particles by lowering the energy 
of the fragmenting jet. Comparing the solid and dashed $\pi^0$ curves indicates 
that the difference could be easily measurable already at $p_T$ = 4 GeV/c. The
lower pair of curves illustrates the small difference expected in the direct 
photon $p_T$ distribution, indicating that another effect of jet quenching will
be to increase $\gamma/\pi^0$.

At CERN, WA98 measured the $\pi^0$ spectrum to nearly 4 GeV/c and did not 
observe any evidence of jet quenching.\cite{wa98pi0} 
However, as discussed above, the soft
physics likely still contributes significantly at this $p_T$, 
masking energy 
loss effects. As the cross section for hard processes will be considerably 
larger at RHIC, the spectra should be measurable to considerably
higher $p_T$. Jet quenching will thus be a very important observable at RHIC.

\section{Prospects for RHIC}

The Relativistic Heavy Ion Collider (RHIC) at Brookhaven National 
Laboratory began operation in summer 2000. RHIC collided 
Au beams at $\sqrt{s}$ = 130 GeV per nucleon, which will be increased
to 200 GeV per nucleon in the next run. RHIC
will also collide smaller nuclei, protons on 
nuclei, and two polarized proton beams at $\sqrt{s}$ up to 500 
GeV. The design luminosity is $2 \times 10^{26} /\rm{cm}^2$/sec for
Au + Au, $10^{31} /\rm{cm}^2$/sec for p + p and $10^{29} /\rm{cm}^2$/sec 
for p + A. Luminosity achieved during the first run reached 10\% of
design value. 

Higher energy and long running time at RHIC will allow in-depth
investigation of the currently tantalizing observables.
With the factor ten increase in center of mass energy,
every collision should be well above the phase transition threshold.
The initial temperature can be expected to significantly exceed
estimates of $T_C$. 
Furthermore, hard processes which provide probes of the early medium
have considerably higher cross sections. Consequently, experiments
will be able to measure $J/\psi$ and other hard processes to higher 
$p_T$ with better statistical significance than before.
Charged and neutral pion spectra at $p_T \ge 5 GeV$ to look for
evidence of jet quenching will be accessible, and the $\gamma/\pi^0$ 
ratio will reach the $p_T$ range where direct
photon yields are calculable.

\subsection{Experiments at RHIC}

To cover the full range of experimental observables,
RHIC has a suite of four experiments. There are 
two large and two small experiments, each optimized differently. 
Together, they form a comprehensive program to fully characterize 
the heavy ion collisions and search for all the 
predicted signatures of deconfinement and chiral symmetry 
restoration. 

Each experiment 
is outfitted with two zero-degree calorimeters of identical design. 
These calorimeters measure neutral particles produced at zero 
degrees, allowing a common method of 
selecting events according to centrality. An event sample with 
interesting behavior observed by one experiment can therefore be 
checked by the other experiments. Many of the hadronic observables 
are measured by two or more of the experiments, so a complete 
picture of the collisions at RHIC will be investigated. 

The two small experiments, PHOBOS and BRAHMS, 
focus on difficult-to-measure regions of rapidity and $p_T$.
PHOBOS is optimized to measure and identify hadrons 
at very low $p_T$ and fits on a (large) table top. The low 
$p_T$ capability provides good sensitivity to formation of 
disoriented chiral condensates. In addition, PHOBOS has a full 
coverage multiplicity measurement, allowing analysis of 
fluctuations and selection of events with unusually numerous 
particles. Particle tracking is done primarily with 
highly granular silicon detectors, allowing very short flight 
paths and minimizing decay of the low $p_T$ hadrons. 
Identification is accomplished by time-of-flight measurements.

BRAHMS maps particle production over a wide 
range of rapidities with good $p_T$ coverage. BRAHMS has two 
movable, small acceptance spectrometers to sample the particle 
distributions; a typical event has only a few particles in each 
spectrometer. Tracking is provided by modest size time projection 
chambers and drift chambers. Particle identification is performed 
via time-of-flight measured by scintillator hodoscopes and via gas 
Cherenkov threshold counters.

The two large experiments each measure many of the predicted QGP 
signatures, along with observables to map the hadronic phase. 
STAR has maximum acceptance for hadrons, allowing 
event-by-event analyses of the final state and reconstruction of 
multi-strange hadron decays. PHENIX 
is optimized for photon and lepton detection and has high rate 
capability and selective triggers to collect statistics on rare 
processes. 

STAR consists of a large acceptance time projection 
chamber, covering full azimuth over two units of rapidity centered 
around mid-rapidity (90 degrees in the laboratory). The TPC sits 
in a solenoidal magnetic field. In the second and third years of RHIC 
running, a silicon vertex tracker and electromagnetic calorimeters 
will be added to improve the efficiency of finding secondary 
vertices and to allow measurement of jets. In the first year, STAR 
had a partial acceptance ring-imaging Cherenkov counter to identify 
a subset of the particles and trigger on high 
$p_T$ hadrons. STAR events include dense information on each of 
many charged particle tracks and are consequently very large; 
approximately one event per second is written.

PHENIX has multiple subsystems to track, identify, 
and trigger on leptons, photons, and hadrons. At midrapidity, 
there is an axial field magnet with two detector sectors, each 
covering 90 degrees in azimuth. Drift, pad, and time 
expansion chambers provide tracking, scintillator 
hodoscope time-of-flight detectors for hadron identification, a 
large ring-imaging Cherenkov counter identifies electrons,
and a highly granular electromagnetic calorimeter is used
for electron and 
photon identification and triggers. Charged particle multiplicity 
and fluctuations are measured with silicon detectors. Forward and 
backward, PHENIX has two cone-shaped magnets outfitted with 
cathode strip detectors for tracking and Iarocci tubes interleaved 
with steel plates for muon identification. The pole tips of the 
central magnet absorb approximately 90\% of the hadrons. PHENIX 
began running with the central arms and 
silicon detectors, with muon measurements commencing in 2001. 

\subsection{First Results}

\begin{figure}
\epsfxsize180pt
\figurebox{}{}{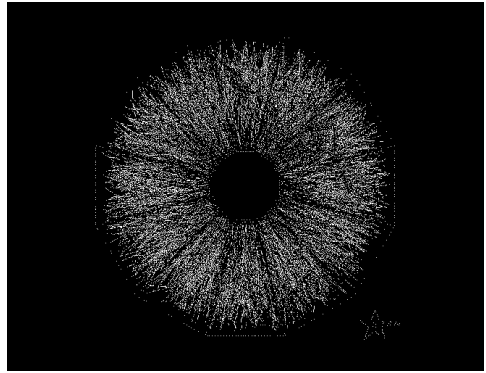}
\caption{Display of a central Au + Au collision at $\sqrt{s}$ =
130 GeV/nucleon in the STAR time projection chamber \protect\cite{star}.}
\label{fig:star}
\end{figure}

Figure 4 shows a central Au + Au collision at $\sqrt{s}$ = 130 
GeV/nucleon, recorded in the STAR TPC.\cite{star} The large number 
of tracks illustrates the challenges for the experiments.
All experiments reconstruct tracks
with good efficiency. STAR has demonstrated successful particle 
identification via dE/dx in these collisions. PHENIX, with excellent
granularity ($\Delta\eta = \Delta\phi = 0.01$) and resolution 
($\approx 8\%/\sqrt{E}$) calorimetry, reconstructs $\pi^0$ and transverse
energy distributions from such events with high particle multiplicity.

The PHOBOS Collaboration has measured the charged particle rapidity
density at midrapidity for the 6\% most central Au + Au collisions.
They find $dN/d\eta = 555 \pm 12 (\rm{stat.}) \pm 35 (\rm{syst.})$ 
at $\sqrt{s}$ = 130 
GeV/nucleon.\cite{phobosmult} The importance of this first measurement 
can be appreciated by looking at the variation in predicted particle
multiplicity for $\sqrt{s}$ = 200 GeV/nucleon in the 
literature\cite{multpred} and for $\sqrt{s}$ = 130 
GeV/nucleon in Figure 5.\cite{multpred} The range
of predictions is almost a factor of 2! Three important factors
control the total number of charged particles produced at midrapidity:
parton multiple scattering (which increases the multiplicity),
nuclear shadowing (the effect is very sensitive to the $x$ and
$Q^2$ dependence), and energy loss in the dense medium (energy
loss tends to increase the number of soft particles at the expense
of $p_T$ in the tail of the distribution). 

\begin{figure}
\epsfxsize=6.5cm
\epsfbox{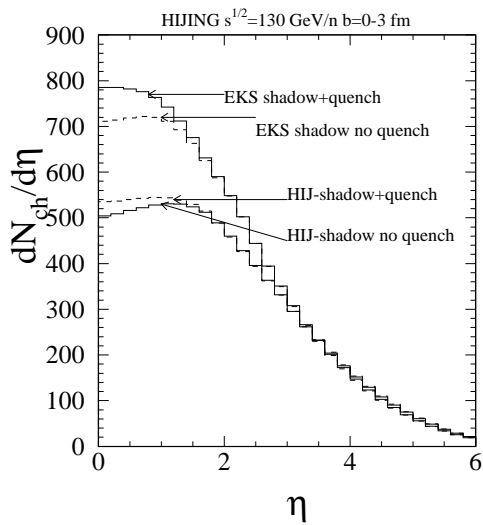}
\caption{Charged particle multiplicity distribution predicted by
HIJING for central Au + Au collisions at $\sqrt{s}$ = 130 GeV/nucleon. 
Different curves correspond to different assumptions of nuclear shadowing 
and parton energy loss \protect\cite{multpred}.}
\label{fig:mult}
\end{figure}

The PHOBOS result shows
excellent agreement with the HIJING model of Wang using quark energy 
loss dE/dx = 1 GeV/fm, gluon dE/dx = 0.5 GeV/fm and nuclear shadowing 
taken from lower $\sqrt{s}$
measurements.\cite{wangjetq} The anti-shadowing prescription of Eskola
and coworkers clearly overpredicts the multiplicity. Of course, it is
difficult to crisply separate three components with a single data
point. The $p_T$ spectrum of hadrons will constrain the parton energy
loss in these collisions; analyses are currently underway. The
shadowing can be determined directly by measurement of hadron yields
at $p_T = 2-6 GeV/c$ in proton-nucleus and proton-proton 
collisions.\cite{wangshadow}
However, at this point we may tentatively conclude that (unless Nature
has conspired to provide some exact cancellations) that nuclear
shadowing appears to saturate and no anti-shadowing occurs.

\section{Conclusions}

I have shown that experiments produce dense interacting matter in the 
laboratory and that we can extract physics from the very complex
interactions between heavy ions. One may ask whether the quark gluon plasma
has been observed in collisions near $\sqrt{s}$ = 20 at CERN, and the answer
must needs be ``probably''. Several predicted signatures have been 
independently measured which defy currently available 
conventional explanations. Correlated 
onset has not been demonstrated, however. The lack of a
coherent theoretical description and the incompleteness of appropriate
dynamic theories make unambiguous conclusions difficult. 
Still missing is experimental determination
of the energy threshold for deconfinement, and characterization of the
properties of the quark gluon plasma state. 

The experimental program in the coming years has its work clearly cut out:
We must determine $T_{init}$ from electromagnetic radiation, measure the
jet quenching and learn to untangle the soft from the hard physics. Observation
of multiple signatures at the same condition will be crucial, and a
measurement of the hadron formation transition would be most helpful. 
RHIC has begun operation, 
and will contribute greatly via an experimental program with common event 
selection to constrain theory via a suite of observables.

\section*{Acknowledgments}
I would like to thank Axel Drees, Xin-Nian Wang, Thomas Ullrich and 
Sam Aronson for valuable discussions and figures for the talk.
This work was 
supportedby the U.S. Department of Energy under grant number 
DE-FG02-96ER40988.

\end{document}